%% file: main.tex
\documentclass[a4paper,usenatbib]{mnras}

\makeatletter

\newcommand{\Rmnum}[1]{\expandafter\@slowromancap\romannumeral #1@}

\newcommand{\gsim}{\lower0.6ex\vbox{\hbox{$\buildrel{\textstyle >}\over{\sim}\ $}}}

\makeatother

\usepackage{newtxtext,newtxmath}
\usepackage{graphicx}	
\usepackage{amsmath}	
\usepackage{multirow}
\usepackage{subcaption}
\usepackage{booktabs}
\usepackage[dvipsnames,svgnames]{xcolor}
\usepackage{comment}
\usepackage{color}

\def\hmpc{h^{-1}{\rm Mpc}}

\def\hkpc{h^{-1}\, {\rm kpc}}

\def\hmsun{{h^{-1} M_{\odot}}}
\def\msun{\, M_{\odot}}
\def\astrid{\texttt{ASTRID} }

\title[Triple and Quadruple Black holes in ASTRID]
{Triple and Quadruple Black Holes in the ASTRID Simulation at $z \sim 2$}

\author[C.~Hoffman et al.]{
Calvin Hoffman,$^{1}$\thanks{E-mail: cjhoffma@andrew.cmu.edu}
Nianyi Chen,$^{1}$
Tiziana Di Matteo,$^{1,2}$
Yueying Ni,$^{1,3}$
Simeon Bird,$^{4}$
\newauthor
Rupert Croft,$^{1,2}$
Abraham Loeb$^{3}$\\
$^{1}$ McWilliams Center for Cosmology, Department of Physics, Carnegie Mellon University, Pittsburgh, PA 15213 \\
$^{2}$ NSF AI Planning Institute for Physics of the Future, 
Carnegie Mellon  University, Pittsburgh, PA 15213, USA \\
$^{3}$Harvard-Smithsonian Center for Astrophysics, 60 Garden Street, Cambridge, MA 02138, USA\\
$^{4}$ Department of Physics \& Astronomy, University of California, Riverside, 900 University Ave., Riverside, CA 92521, USA
}

\date{Accepted XXX. Received YYY; in original form ZZZ}

\pubyear{2023}
\begin{document}
\maketitle

\begin{abstract}
We use the \texttt{ASTRID} cosmological hydrodynamic simulation to investigate the properties and evolution of triple and quadruple Massive Black Hole (MBH) systems at $z = 2-3$.
Only a handful of MBH tuple systems have been detected to date. 
In \texttt{ASTRID}, we find $4\%$ of the $M_{\rm BH}>10^7\,M_\odot$ are in tuples with $\Delta r_{\rm max} < 200\,{\rm kpc}$. 
The tuples systems span a range of separations
with the majority of the observable AGN systems at $\Delta r \sim 50-100$ kpc. They include some of the most massive BHs (up to $10^{10} \msun$) but with at least one of the components of $M_{\rm BH} \sim 10^7 \msun$.
Tuples' host galaxies are typically massive with $M_* \sim 10^{10-11} \msun$. 
We find that $>10\%$ massive halos with $M_{\rm halo} > 10^{13} M_\odot$ host MBH tuples.
Following the subsequent interactions between MBHs in tuples, we found that in $\sim 5\%$ of the triplets all three MBHs merge within a Gyr, and $15\%$ go through one merger.
As a by-product of the complex multi-galaxy interaction of these systems, we also find that up to $\sim 5\%$ of tuples lead to runaway MBHs. 
In \texttt{ASTRID}, virtually all of the ultramassive black holes ($>10^{10} \msun $) have undergone a triple quasar phase while for BHs with
$M_{\rm BH} \sim 10^9 \msun$ this fraction drops to 50\%.
 

\end{abstract}


\input{Sec1_Intro.tex}
\input{Sec2_Method.tex}
\input{Sec3_Result.tex}

\input{Sec4_Conclusion.tex}

\section*{Acknowledgements}

We acknowledge helpful discussions with Tianqing Zhang which lead to the current version of Figure 3.
\texttt{Astrid} was run on the Frontera facility at the Texas Advanced Computing Center.
TDM and RACC acknowledge funding from 
the NSF AI Institute: Physics of the Future, NSF PHY-2020295, 
NASA ATP NNX17AK56G, and NASA ATP 80NSSC18K101. 
TDM acknowledges additional support from  NSF ACI-1614853, NSF AST-1616168, NASA ATP 19-ATP19-0084, and NASA ATP 80NSSC20K0519, and RACC from NSF AST-1909193.
YN acknowledges support from McWilliams Graduate Fellowship and the ITC Postdoctoral Fellowship.
SB acknowledges funding from NASA ATP 80NSSC22K1897.

\section*{Data Availability}

The code to reproduce the simulation is available at \url{https://github.com/MP-Gadget/MP-Gadget}, and continues to be developed. Text file forms of the data presented here and scripts to generate the figures are available. The MBH-multiplet catalog including the host galaxy/halo information at $z=2$ and $z=3$ and the MBH properties (e.g. trajectories, masses, luminosities) are available at a higher time resolution through the ``BH details" files.

\bibliographystyle{mnras}
\bibliography{main.bib}

\end{document}

%% file: Sec1_Intro.tex
\section{Introduction}
The coevolution of galaxies and the supermassive black holes (SMBHs) at their centers via hierarchical galaxy mergers is a key prediction of $\Lambda$CDM cosmology. 
During the galaxy mergers, active galactic nuclei (AGN) can be triggered by the gas driven towards the center of the merger remnant and onto the SMBHs \citep[e.g.][]{DiMatteo2005,Hopkins2008}, making these SMBH pairs observable as dual AGN.
In the most extreme environment of rare density peaks, the halo can host a few massive galaxies due to the infall of satellites into the deep potential well \citep[e.g.][]{Rodriguez-Gomez2015}.
Under such a scenario, it is possible that one or more galaxies together with the MBHs at their centers, fall into the site of an ongoing galaxy merger, and the resulting system would manifest as a multiple AGN if all SMBHs are simultaneously accreting.

Multiple MBH systems are exciting objects to study for a few reasons. 
First, these systems typically reside in the highest density peaks which would later evolve into galaxy clusters in the present day.
These galaxy clusters are the largest virialized objects known to date. 
By studying the multiple AGN system in the early days of the cluster assembly, we learn about the evolution of galaxy clusters, and can thereby potentially reveal new
insights into galaxy evolution.
Second, galaxy mergers are believed to enhance MBH growth and star formation. 
In the case of multiple subsequent galaxy mergers, there could be a phase of continuous increase in the gas supply that spans several hundred Mega years in time.
Both the resulting rapid accretion onto the MBHs and the successive MBH merger could lead to the formation of the most massive SMBHs \citep[e.g.][]{Ni2022b}.
Lastly, multiple MBH systems provide a channel for the accelerated binary formation and coalescences. 
For example, \citet{Bonetti2019} showed that triple-induced coalescences can take up to $30\%$ of the total detected mergers by LISA, and could provide up to 20 LISA events per year even when the standard gas/stars-driven evolutionary channels should fail and MBH binaries were to stall.
This also made multiple MBH systems interesting objects for multi-messenger studies: while previous works have shown that the coalescence time of binary MBHs may be too long for any signatures in the galaxy morphology \citep[e.g.][]{DeGraf2021},
in the case of a triple MBH system, the GW emission would happen when the host galaxies still retain the disrupted morphologies from mergers.

Despite their scientific merit, multiple AGN are difficult to study both observationally and theoretically, mainly because of their scarcity. 
To date, only a handful of triple/quadruple quasar systems have been identified by observations.
At low redshifts, \cite{Pfeifle2019} and \cite{Liu2019} provides compelling evidence for the existence of a triple AGN system SDSS J0849+1114 separated by only a few tens of kpc at $z=0.077$.
The most notable triple quasar systems at high redshifts are QQQJ1432-0106 at $z=2.1$ observed by~\cite{Djorgovski2007}, and QQQ J1519+0627 at $z=1.5$ reported by~\cite{Farina2013}, both on separations of several tens to hundreds of pkpc at $z \sim 2$. 
On the more extreme side, \citet{Hennawi2015} found a quadruple AGN system at $z\sim 2$ embedded in a giant nebula, with one bright quasar and three companion AGNs.


On the theoretical side, studies of multiple MBH systems date back several decades ago, starting with analytical models of many-body interactions and developing into experiments with N-body simulation and derivation of semi-analytical models
\citep[e.g.][]{Volonteri2003, Iwasawa2006, Hoffman2007, Kulkarni2012, Bonetti2016, Bonetti2018a}.
The study of multiple MBHs as observable AGN in a cosmological context and within realistic galactic environments is only made possible recently by very large volume $>100\,{\rm Mpc}^3$ cosmological simulations. 
Among these, \citet{Volonteri2020} studies the host properties of multiple AGN systems up to 6 AGNs in the HorizonAGN simulation.
\citet{Bhowmick2020} focuses on Mpc-scale multiple AGN systems and found that most multiplet systems are expected to be hosted in separate galaxies.

Most of the previous studies, however, either assumed that close MBH triplet (or multiplet) encounters already take place and model the subsequent three-body interaction, or focus on the $\gtrapprox 100\,{\rm kpc}$-scale properties of bright AGN, without detailed modeling of whether these large-scale AGN multiples will evolve into a sub-kpc-scale MBH multiplet system.
In this work, we aim to partially bridge the gap between multiple AGN studies on cosmological scales and triplet MBH encounters on the binary formation scale.
In particular, we will try to address the following questions: what fraction of hundred-kpc scale MBHs can quickly form a close-triplet system, and how many experience ejections/orbit disruptions? During the evolution from the hundred-kpc scale to the sub-kpc scale MBH multiplet systems, how many and what kinds of these systems power multiple AGN?

This paper is organized as follows: in Section \ref{sec:sim}, we introduce the \texttt{ASTRID} simulation, in particular the MBH modeling, and describe our selection criterion for the triple and quadruple MBH systems from the simulation;
in Section \ref{sec:results}, we present our major results including the MBH masses, luminosities, separations, and AGN observability of our MBH tuples. We also trace the evolution of the MBHs tuples into mergers and runaway MBHs. Finally, we investigate the role of MBH triplets in the formation of SMBHs.

%% file: Sec2_Method.tex
\section{Simulation}
\label{sec:sim}
\begin{figure*}
    \centering
    \includegraphics[width=\textwidth]{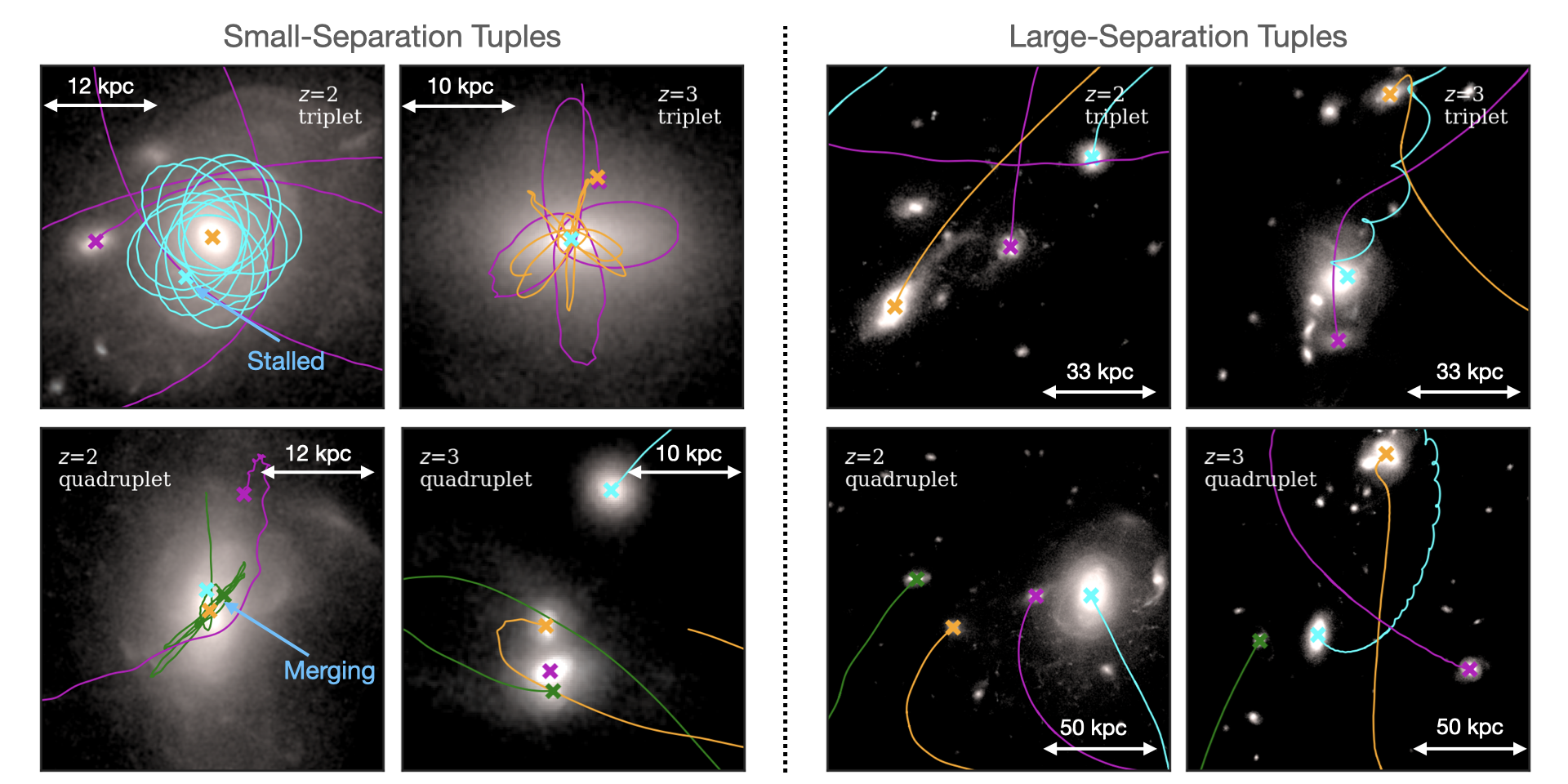}
    \caption{Selected triple and quadruple systems of 30 kpc separation and 200 kpc separation at $z = 2$ and $z = 3$ plotted on top of their host galaxies. 
    Cross markers indicate MBHs and their trajectories are plotted in the same color. 
    The top right corner of the images displays the redshift in which the image is generated.
    The small-separation tuples are found in ongoing galaxy mergers, with frequent tidal disruption of host galaxies. 
    The large-separation tuples are embedded in their own hosts.
    For small-separation tuples, we plot the BH trajectories with respect to the trajectory of the most massive MBH in the tuple.
    For large separation galaxies, we plot the trajectories with respect to the center of the current frame.
    }
    \label{fig:tuple_images}
\end{figure*}

\begin{figure*}
\centering
  \includegraphics[width=0.97\textwidth]{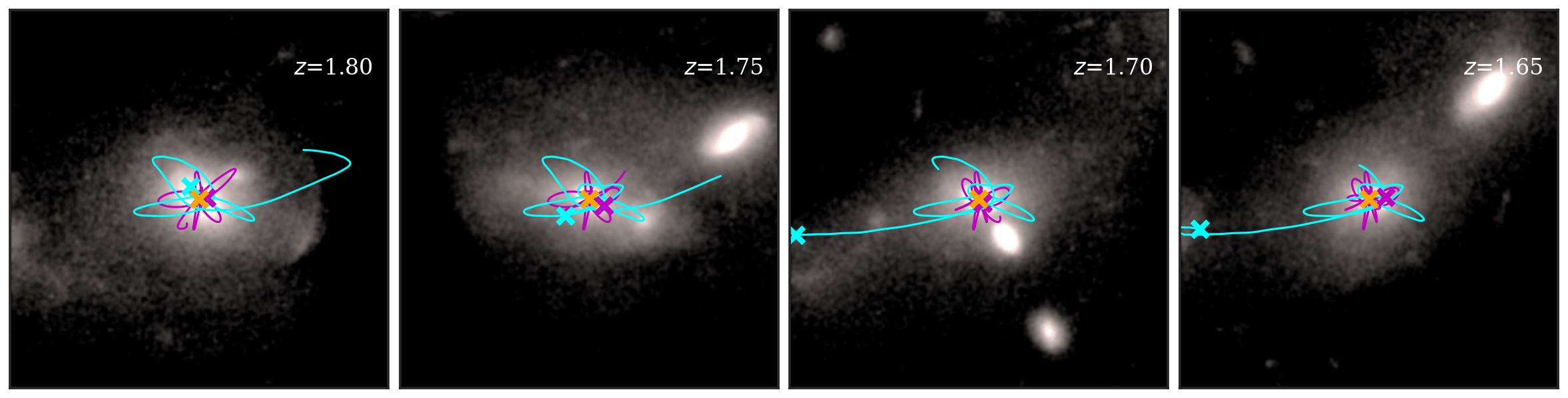}
  
  \caption{The subsequent evolution of a $z=2$ triple MBH system leading to a potential runaway BH. The \textit{cyan} BH has already fallen in the merged galaxy at $z=1.8$, but it was ejected at $z=1.7$ to the outskirts of the host galaxy when another incoming dense galaxy gets into its orbit.}
\label{fig:z3_dr30_triple_evolution}
\end{figure*}

\astrid is a cosmological hydrodynamical simulation performed using a new version of the \texttt{MP-Gadget} simulation code.
The simulation comprises $5500^3$ cold dark matter (DM) particles in a $250 \hmpc$ side box, with an equal number of SPH hydrodynamic mass elements present at the outset.
The cosmological parameters utilized in \astrid were obtained from \cite{Planck}, with values of $\Omega_0=0.3089$, $\Omega_\Lambda=0.6911$, $\Omega_{\rm b}=0.0486$, $\sigma_8=0.82$, $h=0.6774$, $A_s = 2.142 \times 10^{-9}$, $n_s=0.9667$.
The gravitational softening length for both DM and gas particles is $\epsilon_{\rm g} = 1.5 \hkpc$.
\astrid achieves a dark matter particle mass resolution of $9.6\times 10^6 \msun$ and $M_{\rm gas} = 1.3 \times 10^6 \msun$ in the initial conditions.
The simulation has reached $z=1.5$ with plans to reach $z=1$.

In order to account for the physics involved in the creation of galaxies and SMBHs, as well as the feedback mechanisms associated with supernovae and active galactic nuclei (AGN), \astrid employs several sub-grid models.
Additionally, the simulation factors in effects such as inhomogeneous hydrogen and helium reionization, and the influence of massive neutrinos.
A summary of the physical models utilized in the simulation is provided below, with a more comprehensive explanation available in the introductory paper \cite{Ni2022,Bird2022}.

In \astrid, gas is allowed to cool via primordial radiative cooling \citep{Katz1996ApJS..105...19K} and via metal line cooling, with the gas and stellar metallicities traced following \cite{Vogelsberger:2014}.
Patchy reionization of hydrogen is achieved via a semi-analytic method based on radiative transfer simulations \citep{Battaglia2013ApJ...776...81B}, utilizing a spatially varying ultra-violet background.
The ionizing ultra-violet background from \cite{FG2020} is employed for ionized regions, with gas self-shielding being factored in as outlined in \cite{Rahmati2013}.
Star formation in \astrid is based on a multi-phase model for stellar formation as described in \cite{SH03}, accounting for the influence of molecular hydrogen~\citep{Krumholz2011ApJ...729...36K}.
Type II supernova wind feedback is incorporated into the simulation in accordance with \cite{Okamoto2010}, with wind speeds proportional to the local one-dimensional dark matter velocity dispersion.
 
Subgrid models for SMBH applied in \astrid can be summarized as follows.
SMBHs are represented by particles that can accrete gas, merge and apply feedback to their baryonic surroundings.
Halo-based seeding is used where the BHs are seeded in halos with $M_{\rm halo,FOF} > 5 \times 10^9 \hmsun$ and $M_{\rm *,FOF} > 2 \times 10^6 \hmsun$. 
Seed masses are stochastically drawn from a power-law probability distribution, with a mass between $3\times10^{4} \hmsun$ and $3\times10^{5} \hmsun$ and power-law index $n = -1$.

A Bondi-Hoyle-Lyttleton-like prescription \citep{DSH2005} is used to estimate the gas accretion rate onto the BH:
\begin{equation}
\centering
\label{equation:Bondi}
    \dot{M}_{\rm B} = \frac{4 \pi \alpha G^2 M_{\rm BH}^2 \rho}{(c^2_s+v_{\rm rel}^2)^{3/2}}
\end{equation}
where $c_s$ and $\rho$ are the local sound speed and density of gas, $v_{\rm rel}$ is the relative velocity of the BH with respect to the nearby gas and $\alpha = 100$ is a dimensionless fudge parameter to account for the underestimation of the accretion rate due to the unresolved cold and hot phase of the subgrid interstellar medium in the surrounding.
We allow for short periods of super-Eddington accretion in the simulation but limit the accretion rate to two times the Eddington accretion rate.
The BH radiates with a bolometric luminosity $L_{\rm bol}$ proportional to the accretion rate $\dot{M}_\bullet$, with a mass-to-energy conversion efficiency $\eta=0.1$ in an accretion disk according to \cite{Shakura1973}.
\begin{equation}
\centering
\label{equation:Lbol}
    L_{\rm Bol} = \eta \dot{M}_{\rm BH} c^2
\end{equation}
$5\%$ of the radiated energy is coupled to the surrounding gas as the AGN feedback.
The feedback from SMBHs includes what is often referred to as quasar-mode or thermal feedback, as well as kinetic feedback.

The dynamics of the SMBHs are modeled with a newly developed (sub-grid) dynamical friction model \citep{Tremmel2015,Chen2021} to replace the original implementation that directly repositioned the BHs to the minimum local potential.
This new model provides an improved treatment for calculating BH trajectories and velocities.
However, BH particles near the host's center are subject to spurious dynamical heating from other particles significantly larger than the galactic component they represent.
This effect may reflect in the initial stages of the BH's orbit when it resides in the central region of the host.
Two BHs merge if their separation is within two times the spatial resolution $2\epsilon_g$, once their kinetic energy is dissipated by dynamical friction, and they are gravitationally bound to each other.
To reduce the noisy gravitational forces (dynamical heating) acting on the small seed mass black holes, another BH mass tracer, the dynamical mass $M_{\rm dyn}$, is used to account for the force calculation of BH, including the gravitational force and dynamical friction. When a new BH is seeded, the corresponding $M_{\rm dyn} = M_{\rm dyn,seed} = 10^7 \hmsun$, which is about $1.5 M_{\rm DM}$.
This approach alleviates dynamic heating and stabilizes the BH motion in the early growth phase.
$M_{\rm dyn}$ is kept at its seeding value until $M_{\rm BH}>M_{\rm dyn,seed}$. After that, $M_{\rm dyn}$ grows following the BH mass accretion.
The validation of the dynamical friction model in cosmological simulations is described in \cite{Chen2021}.

\astrid provides ``BH details'' files that log the BH evolution at high time resolution and allow detailed analysis of the BH orbital trajectories and light curves. 
Among the BH snapshots used in our analysis, the median time spacing is $\sim 4.8 \times 10^{-5} \rm \, Gyr$.

\begin{figure*}
\centering
\includegraphics[width=0.99\textwidth]{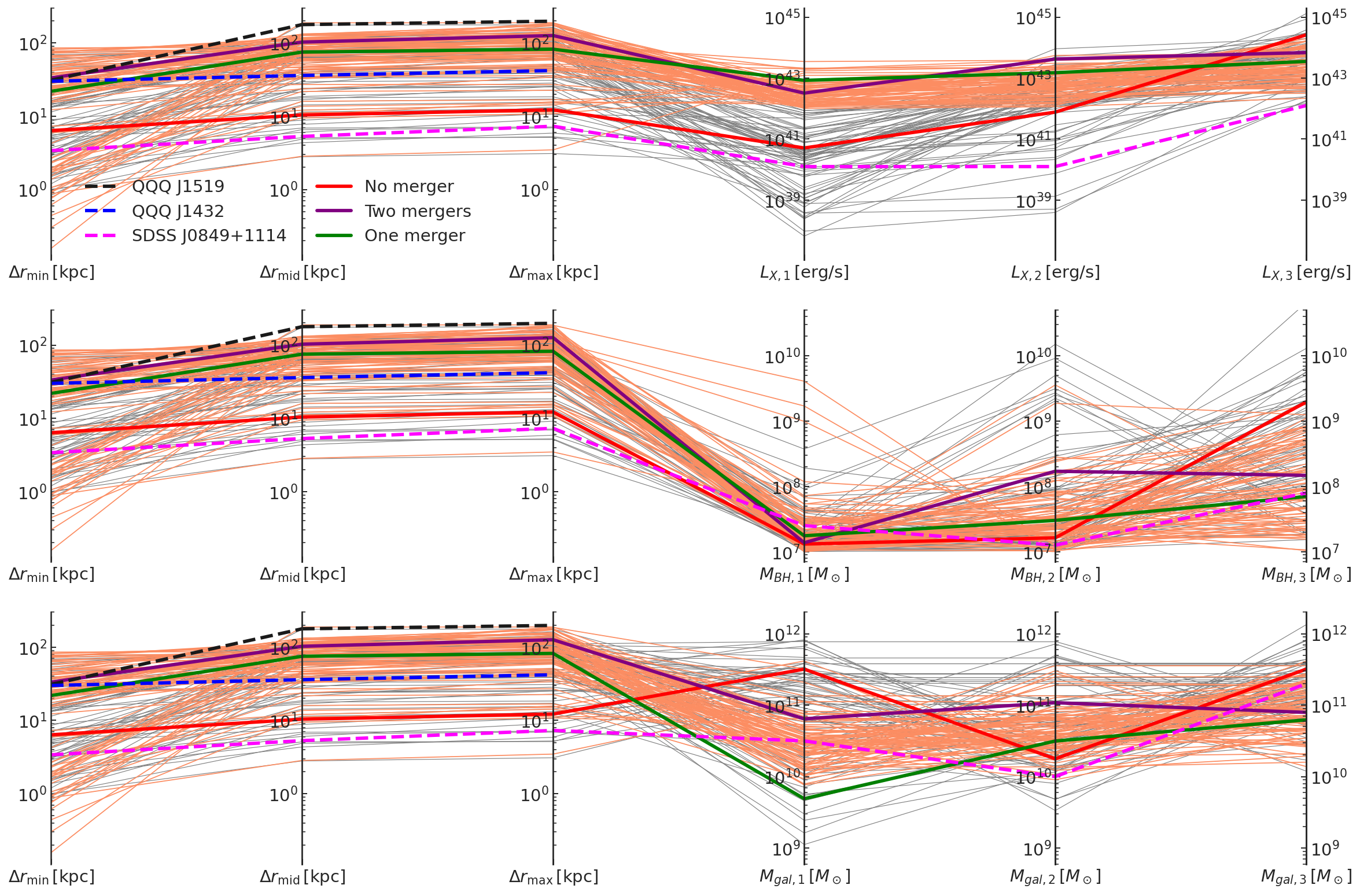}
  \caption{Properties of triple MBH systems including the pairwise separations (\textbf{all rows, left half}), X-ray luminosities (\textbf{top row, right half}), black hole masses (\textbf{middle row, right half}), and host galaxy masses (\textbf{bottom row, right half}). 
  Each line connects all properties of a single triplet system.
  We plot a random sample of 150 triplets in our catalog (\textbf{\textit{thin lines}}), where the systems with all three MBHs above $L_X = 10^{42}\,{\rm erg/s}$ are shown in \textbf{\textit{orange}} and the rest in \textbf{\textit{grey}}.
  We also show selected examples of a no-merger triplet (\textbf{\textit{red}}), a two-merger triplet (\textbf{\textit{purple}}), and a one-merger triplet (\textbf{\textit{green}}).
  For comparison, we also plot the available properties of three observed systems QQQ J1519 (\textbf{\textit{black}}), QQQ J1432 (\textbf{\textit{blue}}), and SDSS J0849 (\textit{\textbf{magenta}}) in \textbf{\textit{dashed lines}}.
  }
\label{fig:lx_mbh_mgal}
\end{figure*}

\subsection{Triple and Quadruple Black Hole Systems Selection}
\label{subsec:criterion}

Among all MBHs in the simulation at a fixed redshift, we define an MBH tuple as $n-$many MBHs with the maximum pairwise separation $\Delta r_{\rm max} < 200\rm {\rm kpc}$.
This separation criterion is motivated by the typical maximum distances at which high-redshift triple quasars have been observed.
Out of the $\Delta r_{\rm max} < 200\rm {\rm kpc}$ tuples, we further select a sub-sample of small-separation tuples at $\Delta r_{\rm max} < 30\rm {\rm kpc}$ to match the distances of the observations by \cite{Djorgovski2007} and \cite{Liu2019}, which are typically MBHs in an ongoing galaxy merger.

For this work, we only focus on the massive end of our population by restricting to MBH tuples with all MBHs above $10^7\,M_\odot$.
To avoid double counting in the triplet and quadruplet samples, we select our MBH tuples in a top-down fashion following \cite{Volonteri2022}: we look for the $\Delta r_{\rm max} < 200\rm {\rm kpc}$ tuples with the most MBHs and remove them from our MBH catalog (which is 7 in our case); then we go to tuples with one fewer MBH and remove those, until we approach 4-MBH and 3-MBH tuples.

After the triple and quadruple selection procedure described above, we find a total of 588 (174) MBH triplets(quadruplets) at $z=2$; 97 (28) MBH triplets(quadruplets) at $z=3$ with $\Delta r_{\rm max} < 200\rm {\rm kpc}$.
Out of these, 
174 (24) MBH triplets (quadruplets) have $\Delta r_{\rm max} < 30\rm {\rm kpc}$ at $z=2$, and 37 (4) MBH triplets(quadruplets) have $\Delta r_{\rm max} < 30\rm {\rm kpc}$ at $z=3$.

We identify the host galaxies of the MBHs with \texttt{Subfind} \citep{Springel2001}, but note that during the close encounters of galaxies, \texttt{Subfind} may not be able to separate the merging systems well. 
Finally, when tracing the MBH and galaxy properties back in redshift, we always follow the more massive progenitor if the MBH of interest has gone through prior mergers.

Figure \ref{fig:tuple_images} shows the images of a MBH tuple in each catagory: small-separation/large-separation tuples, $z=2$ and $z=3$ tuples, triples and quadruples.
Cross markers represent MBHs in the current frame and their trajectories in the past $\sim 300\,{\rm Myrs}$ are displayed in the respective color.
The small-separation tuples are found in ongoing galaxy mergers, with frequent tidal disruption of host galaxies and soon-to-be MBH mergers.
The large separation tuples are mostly embedded in their own hosts.
Unique to the small-separation tuples are frequent cases of MBHs being significantly off-center.

 Figure \ref{fig:z3_dr30_triple_evolution} depicts the subsequent evolution of a selected small-separation MBH triplet from $z=1.8$ to $z=1.65$ in redshift increments of $0.05$. 
 This system is in our $\Delta r<30\,{\rm kpc}$ triplet catalog at $z=2$, and the host galaxies of the three MBHs already merged at $z=1.8$.
 The orbital size of the two smaller MBHs fell within $10\,{\rm kpc}$ of the galaxy center.
 However, we see that between $z=1.75$ and $z=1.7$, the orbit of the cyan MBH is strongly disrupted, likely due to the interaction with another dense incoming galaxy.
 The cyan MBH is ejected to the outskirt of the host and is a viable runaway MBH candidate.
 From the background host galaxy images, we can see signatures of tidal disruption of the host galaxy during the triplet encounter.
 However, we note that \astrid does not explicitly model the possible nuclear stellar clusters around the MBHs, which may be bound to the MBHs and do not get completely stripped during the dynamical interactions.
 We are working towards a more delicate dynamical model around MBHs in our simulations.
 

%% file: Sec3_Result.tex
\section{Results}
\label{sec:results}
\subsection{Overview of the MBH Tuple Population}
\label{sec:overview_properties}

\begin{figure}
\centering
  \includegraphics[width=0.49\textwidth]{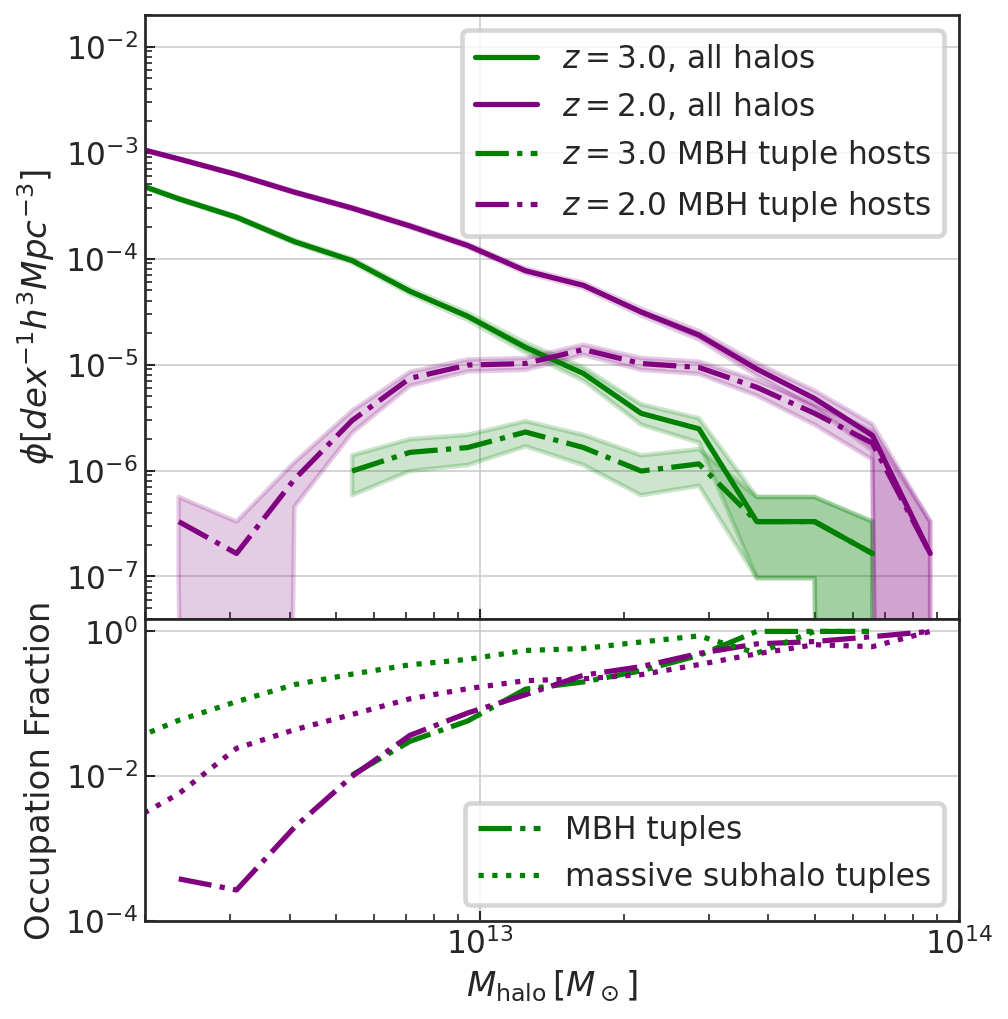}
  \caption{\textbf{Top:} The massive end of the \texttt{ASTRID} halo mass function at $z=2$ (\textit{purple solid}) and $z=3$ (\textit{green solid}), compared with the mass function of massive BH tuple hosts (\textit{dot-dashed}). \textbf{Bottom:} the fraction of halos hosting MBH tuples as a function of halo mass. The trend is very similar for $z=2$ and $z=3$, and $>10\%$ of halos with $M_{\rm halo} > 10^{13}\,M_\odot$ host more than three $M_{\rm BH} > 10^7\,M_\odot$ MBHs. For comparison, we also plot the fraction of halos hosting massive subhalo tuples (\textit{dotted}).}
  \label{fig:z2_z3_hmf}
\end{figure}

We begin with an overview of the properties of the triple MBH systems in our catalog at $z=2$ and $z=3$.
In Figure \ref{fig:lx_mbh_mgal}, we show the characteristic properties of our triple systems, including pairwise separations, X-ray luminosities, black hole masses, and host galaxy masses.
Each system is represented by one line in each row, which connects all of the properties above (for example, for the red system we can read off its minimum pairwise separation to be $\Delta r_{\rm min}\sim 5\,{\rm kpc}$ from the left-most axis).
Here the subscript ``1" always represents the faintest BH and ``3" the brightest (e.g. $M_{\rm BH, 3}$ is the mass of the brightest BH but not necessarily the heaviest one).

Looking at the entire MBH triplet sample (all thin lines), we can see that the minimum pairwise separation spans across a wide range from 0.1 to 200 kpc, while the other two sides of the triangle are usually large (mostly above $20\,{\rm kpc}$).
We also show the pairwise separation of two observed high-redshift triple quasars \citep[][]{Djorgovski2007, Farina2013} and a low-redshift system \citep[][]{Pfeifle2019, Liu2019}, we can see that both systems fall in the typical separation range of our triplet samples.

In Figure \ref{fig:lx_mbh_mgal}, we highlight systems with all three MBHs bright ($L_X > 10^{42}\,{\rm erg/s}$, dubbed ``3-AGN systems" in this work) in orange.
Comparing these bright triplets with the background grey population, we can see that they are mostly characterized by three similar-mass BHs (middle row), and with galaxy masses falling in the mid-range among all triple host galaxies (bottom row).
These 3-AGN systems are usually embedded in at least two host galaxies instead of a single merged galaxy: from the third/fourth columns of the bottom panel, we see that $M_{\rm gal, 1}$ and $M_{\rm gal, 2}$ are different for most orange lines.
A lot of the faint MBHs are found in post-merger large galaxies with $M_{\rm gal, 1} > 10^{11}\,M_\odot$ (the thin lines shooting out from this range are mostly grey).

Figure \ref{fig:z2_z3_hmf} compares the massive end of the \texttt{ASTRID} halo mass function (HMF) at $z=2$ and $z=3$ with the HMF of MBH tuple hosts in the top panel and shows the fraction of halos hosting MBH tuples as a function of halo mass in the bottom panel.
Comparing the HMFs to the mass function of MBH tuple hosts, we observe that as halo mass increases, the tuple-hosting halo tends towards its respective HMF.
This is confirmed by the bottom graph, as the fraction of halos hosting MBH tuples reaches $100\%$ at $\sim 4 \times 10^{13}\ M_\odot$ ($z=3$) and $\sim 9 \times 10^{13} M_\odot$ ($z=2$).
For extremely massive halos, there should exist at least one of the classifications of an MBH tuple.

In previous works, a typical semi-analytical method to study the evolution of multiple MBHs in a host halo was to first establish a connection between the mass distribution of satellite halos within a massive halo, and then determine the MBH in each satellite by a $M_{\rm halo}-M_{\rm BH}$ scaling relation \citep[e.g.][]{Hoffman2007, Kulkarni2012}.
However, there may be a large scatter in both the $M_{\rm halo}-M_{\rm gal}$ as well as the $M_{\rm gal}-M_{\rm BH}$ relation.
To verify the representation of MBHs by their host subhalos, in Figure \ref{fig:z2_z3_hmf} we also include the occupation fraction of subhalo tuples that supposedly should host $M_{\rm BH} > 10^7\,M_\odot$ MBHs.
Here we use the $M_{\rm halo}-M_{\rm BH}$ relation in Equation 29 of \cite{Kulkarni2012}: the corresponding subhalo masses for a $M_{\rm BH} > 10^7\,M_\odot$ BH are $M_{\rm halo} > 2.85\times 10^{11}\,M_\odot$ and $M_{\rm halo} > 1.87\times 10^{11}\,M_\odot$ at $z=2$ and $z=3$, respectively.
We see that at the high-mass end ($M_{\rm halo} > 10^{13}\,M_\odot$), the MBH tuples are well-represented by the subhalo tuples, but for smaller halos, the subhalo tuples could result in an over-estimation of the MBH tuples.


\subsection{Multiple AGN Observability}
\label{sec:agn_offsets_and_observability}

\begin{figure*}
\centering
  \includegraphics[width=0.99\textwidth]{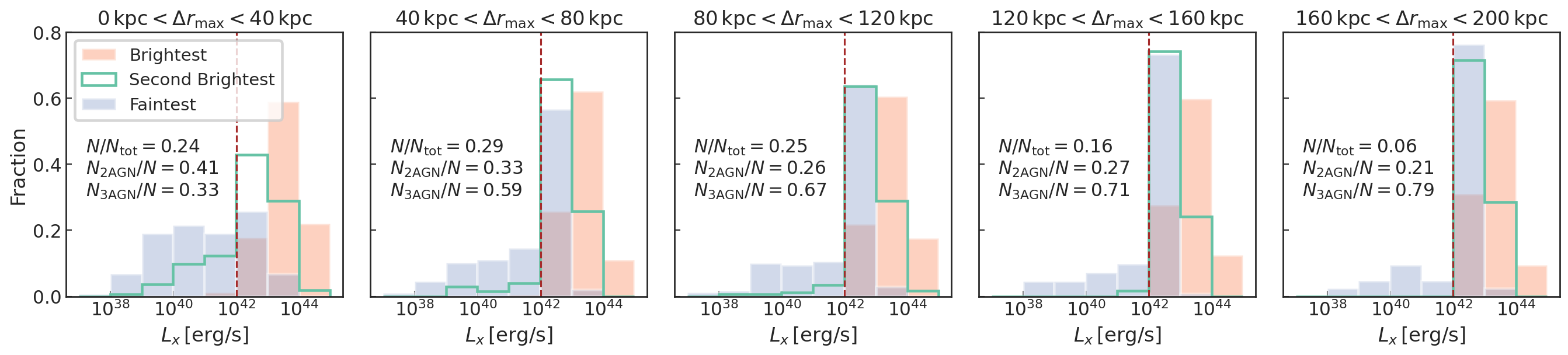}
  \caption{X-ray luminosity distribution of the brightest (\textit{orange}), second brightest (\textit{green}), and the faintest (\textit{purple}) MBH in the triples, for the triples at different maximum separations.
  In each panel, ``$N_{\rm tot}$" denotes the total number of MBH triples;
  ``$N$" is the number of MBH triples in this separation bin; ``$N_{\rm 2AGN}$" is the number of 2-AGN triples at this separation; ``$N_{\rm 3AGN}$" is the number of 3-AGN triples at this separation.
 }
  \label{fig:lx_rbin}
\end{figure*}

\begin{figure}
\centering
\includegraphics[width=0.48\textwidth]{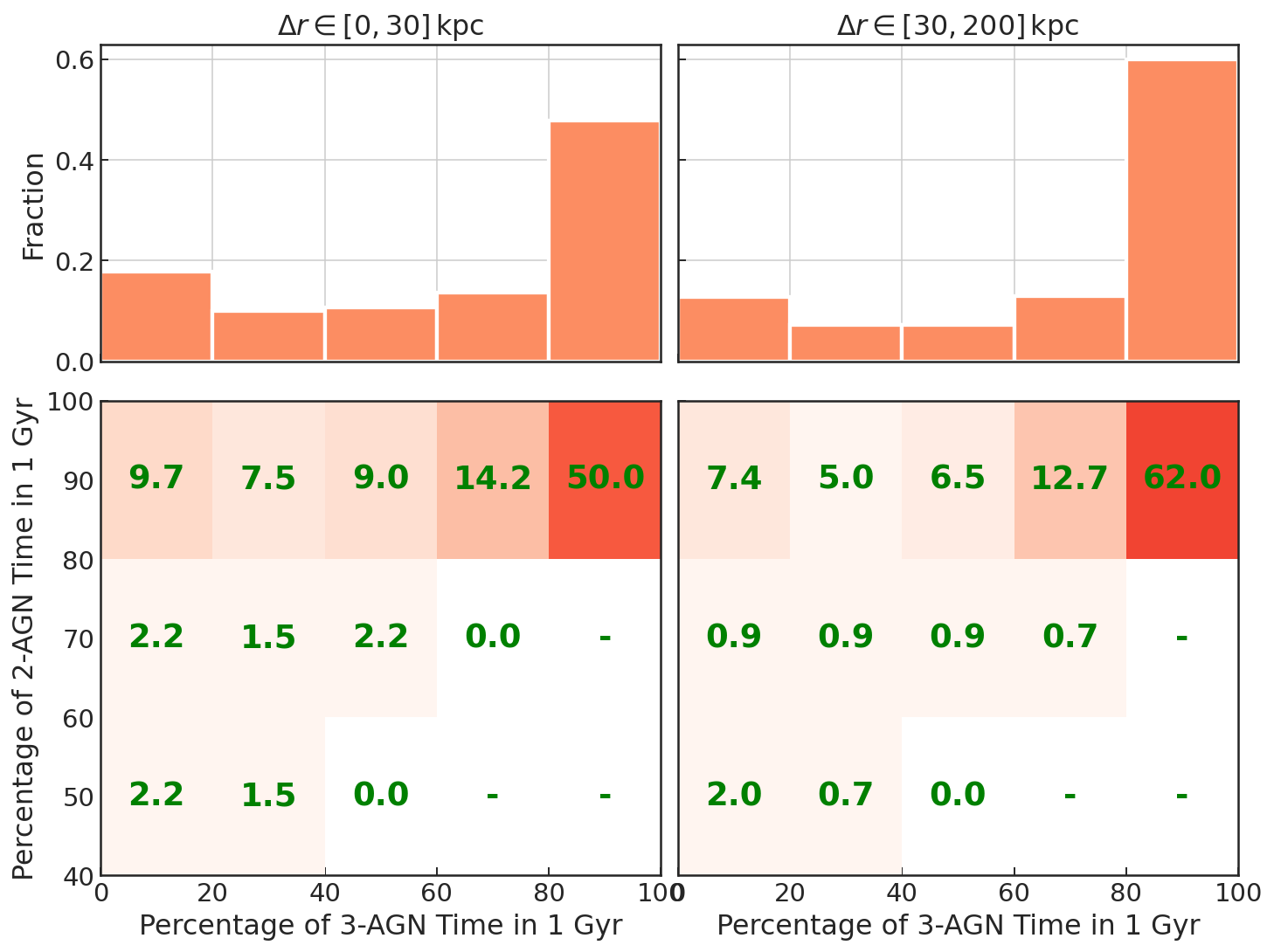}
  \caption{\textbf{Top:} distribution of the fraction of time that all three MBHs are simultaneously bright for $1\,{\rm Gyr}$ before $z=2$ or $z=3$. \textbf{Bottom:} a 2D histogram showing the fraction of 3-AGN time verses 2-AGN time of the triplet across 1 Gyr. The number labels the percentage of triplets in each bin.}
  \label{fig:triples_z3_lumx_perc}
\end{figure}

In the previous section, we have established that $>10\%$ of massive halos with $M_{\rm halo} > 10^{13}\,{M_\odot}$ host MBH tuple. 
However, those multiple massive BHs cannot always be observed as AGN tuples, as not all individual BHs in the tuple systems are simultaneously powered by active gas accretion. 
Some BHs can be stripped of gas supply and become inactive due to the tidal disruptions during galaxy mergers.
It is also possible that the largest MBH accretes dominantly and heats so much gas that it either uses up or evaporates the cold gas reservoir of the smaller BH.
Indeed, in Figure \ref{fig:lx_mbh_mgal}
we have seen that in many of the MBH triplet examples, one or more of them have X-ray luminosities below $10^{42}\,erg/s$.
In this section, we investigate the observability of MBH tuples as AGN tuples.

In Figure \ref{fig:lx_rbin}, we show the distribution of the X-ray luminosity of the three MBHs in the triplet at different maximum separations.
We see that the majority ($\sim 80\%$) of the MBH triplets are within a maximum separation of $120\,{\rm kpc}$ of each other, while only $\sim 20\%$ are separated by more than $120\,{\rm kpc}$.
However, the close separation triples are involved in galaxy mergers and interactions (e.g. see the example in Figure \ref{fig:tuple_images}), so it is more likely for MBHs in close encounters to be deactivated.
From the left-most panel of Figure \ref{fig:lx_rbin}, we can see that in $\sim 70\%$ close-separation triples, the AGN activity of the third MBH is quite low relative to its mass.
The second-brightest MBH at $\rm \Delta r_{\rm max} < 40\,{\rm kpc}$ is also more likely to be inactive compared with other separation bins.
By comparing across all panels of Figure \ref{fig:lx_rbin}, we can see that the chances for all three MBH to be active increase monotonically with their maximum separation, and $>30\%$ of the triple MBH at $\Delta r_{\rm max} < 80\,{\rm kpc}$ may only be seen as duals.
A 3-AGN system will most likely have a maximum separation of $40-120 \,{\rm kpc}$, when combining both the triple MBH separation distribution and the chances of all three MBH being active.

Up until now, our multiple MBH sample only shows one snapshot of AGN activity during the evolution of the system.
However, the tidal stripping takes effect gradually, and there is also periodicity in the MBH accretion.
We now investigate the light curves of those MBH tuples and estimate the potential observability of MBH tuples as AGN tuples in a one-Gyr window.

\begin{figure*}
\centering
  \includegraphics[width=\textwidth]{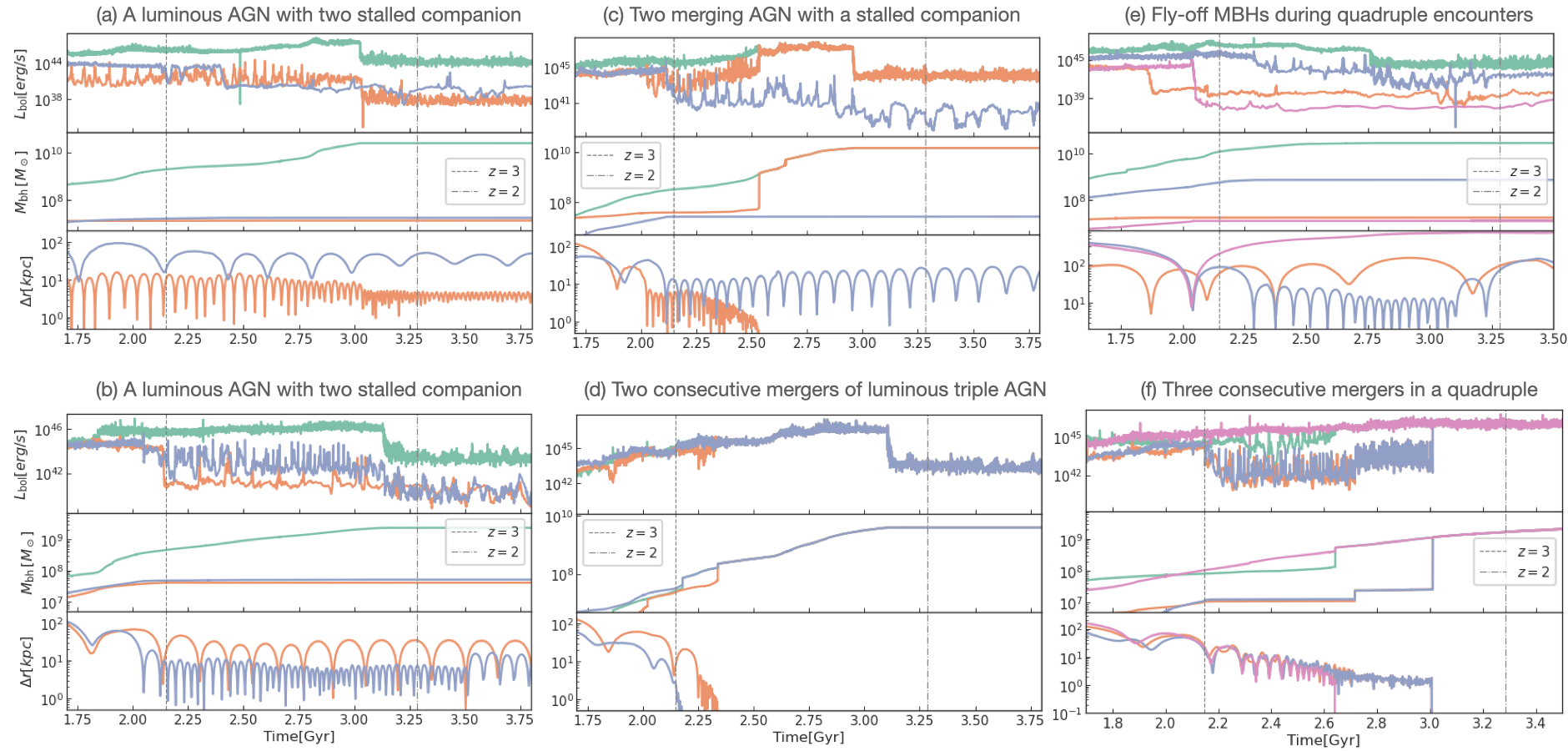}
  \caption{Examples of the triple and quadruple MBHs with diverse evolution paths.  \textbf{(a,b):} a very massive ($M_{\rm BH} > 10^9\,M_\odot$) with two small companions ($M_{\rm BH} \sim 10^7 \,M_\odot$), with both companions stalling after the galaxy mergers and becoming inactive. \textbf{(c)}: Two of the three MBHs in the triplet merged within a few Myrs of their encounter, while a third MBH entering at a similar time got tidally stripped and stalled. \textbf{(d)}: Three similar-mass and luminosity AGN go through quick and consecutive mergers within $<500\,{\rm Myrs}$ of their encounter. \textbf{(e)}: A MBH quadruplet with very unequal masses. The two least massive MBHs experience orbital disruptions during the encounter with the larger MBHs.
  \textbf{(f)}: three consecutive mergers among an MBH quadruple. Two pairs of similar mass MBHs merge first, followed by another merger between the two remnants.
The most massive MBH among the tuple (or the most massive remnant in case of mergers) experiences rapid growth in mass by an order of magnitude during the tuple phase. }
  \label{fig:histories}
\end{figure*}

Figure \ref{fig:triples_z3_lumx_perc} illustrates the probability that any two and all three of the MBHs in the $z=3$ (and $z=2$) tuples are considered as AGN (with $L_X > 10^{42} {\rm [erg/s]}$) during their evolution in the past Gyr.
The $x$ (and $y$) axis in the lower panels gives the percentage of time that an MBH triplet system is considered as 3-AGN (2-AGN). 
The value of the 2D histogram gives the percentage of the MBH triple systems that lies within the given xy bin.
Focusing on the highest time fraction bin ($>80$\%), we can see that among all the MBH triples ($\Delta r_{30}$ and $\Delta r_{200}$) at $z=3$, more than 90\% percent of them can be observed to have at least 2 AGN during for most of the time evolution in the past Gyr, and more than 50\% can be observed as a 3-AGN system.
Thus, even in a 1 Gyr time range, the 3-AGN observability is relatively high (but maybe lower if we increase the luminosity threshold for high-redshift observabilities).
 About half of the MBH triple system might be missed with 1 under-luminous component, likely due to the tidal disruptions during galaxy interaction and merger.

\subsection{MBH Tuples to Mergers, Wanderers, and Ejections}
\label{sec:mergers_ejections}

\subsubsection{A Case Study}

MBH tuples at tens to hundreds of kpc separations have a large variety of fates, due to the complex interactions among themselves, among their hosts as well as with other incoming satellites.
In our MBH triple and quadruple samples, we can find examples of both quick mergers between the three MBHs in the triple, as well as orbital disruption or stalling between an MBH pair due to the entrance of the third.
Each scenario also leads to different AGN luminosities and MBH mass growth.
In this subsection, we demonstrate the possible evolution paths of MBH tuples with a few representative examples.
Figure \ref{fig:histories} displays the luminosity (\textit{top}), BH mass (\textit{middle}), and pairwise separation (\textit{bottom}) evolutions for a collection of tuple systems.

\textbf{A bright quasar with stalled companions}:
Among the examples, we find that tuple-induced stalled orbits are typically found in systems involving minor mergers with $q<0.01$, and with the dominant growth of a central SMBH. 
Panels (a) and (b) of Figure \ref{fig:histories} depict a luminous AGN with two stalled companions.
These two stalled MBHs are the more massive counterparts of the wandering BHs previously studied in works such as \cite{Bellovary2019}, \cite{Ricarte2021}, and \cite{DiMatteo2022}.
In both cases, the infalls of two $\sim 10^{7}\,M_\odot$ both leads to the dominant growth of the central SMBH and the deactivation of the infalling MBHs, leaving the smaller MBHs as potential wanderers with very insignificant mass growth in the following two Gyrs.

Panel (c) in Figure \ref{fig:histories} shows two merging AGN with a stalled companion.
In this case, the more massive MBHs have a smaller mass contrast at the beginning of the triplet phase, leading to an in-simulation merger within $\sim 300\,{\rm Myrs}$.
This is followed by a suppression of growth in the lightest BH, which ends up wandering on a $\sim 10\,{\rm kpc}$ orbit.

\textbf{Quick, consecutive mergers}:
Panels (d) and (f) show two consecutive mergers of luminous triple AGN and three consecutive mergers in a quadruple MBH, respectively.
In the case of the triple, the three MBH comes into the triplet with similar masses and luminosities, and are all on the smaller end of the mass distribution but are accreting at relatively high rates.
The system undergoes two consecutive mergers in a relatively short period ($\sim 0.35\ \rm Gyr$ between the first and second merger).
The merger remnant then goes through a phase of fast mass growth in the following $200\,{\rm Myrs}$, during which it becomes ten times heavier.

In the case of the quadruple system, the merger first happens within two pairs of BHs each with very similar masses and luminosities.
Then the two merger remnants also merged.
We also see some level of suppressed growth of the small pair, but all four MBHs are above the $L_x = {10^{42}}\,{\rm erg/s}$ threshold most of the time.
In previous work by \cite{Ni2022b}, it has been shown that triple MBH mergers could result in an extremely massive BH with $M_{\rm BH}\sim 10^{11} M_\odot$.
In our example of a triplet merger, the remnant also outgrows all of its progenitors by two orders of magnitude.
This example again indicates the contribution of MBH tuples to the formation of the largest SMBHs. 
We will come back to this point in the following section.

\begin{figure*}
\centering
  \includegraphics[width=0.93\textwidth]{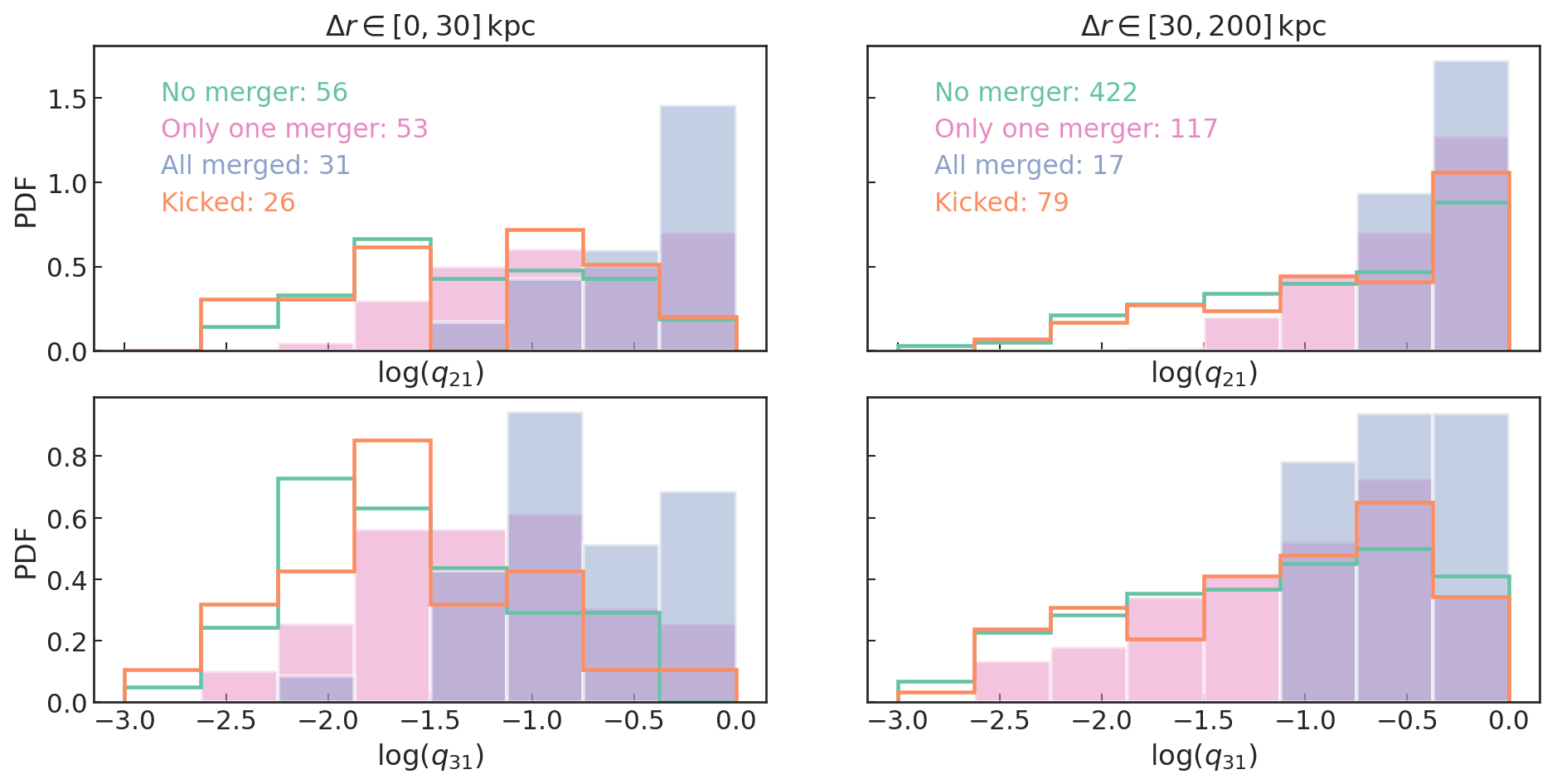}
  \caption{\textbf{Top:} the mass ratio between the two most massive MBHs in the triplet, where the close-separation triplets are shown on the \textit{left}, and the large-separation triplets are shown on the \textit{right}. We classify the triplets into no-merger systems (\textit{green}), one-merger systems (\textit{pink}), and systems in which all three MBHs have merged (\textit{purple}). We also show a category in which there is a possible ``kick" to one of the MBHs causing a sudden orbital enlargement in the triplet phase (\textit{orange}).
  \textbf{Bottom:} the mass ratio between the lightest and heaviest MBH in the triplets. All large-separation systems that have gone through two consecutive mergers consist of three similar-mass MBHs with pairwise mass ratios above $0.1$.} 
  \label{fig:mergers}
\end{figure*}

\textbf{MBH ejections on galactic-scales}:
Very recently, \cite{vanDokkum2023} found signatures of a wake of shocked gas and young stars, where the line ratios, colors, and overall morphology are consistent with an ejected SMBH moving through the circumgalactic medium at high speed.
One mechanism that could produce such an ejection is the interaction between three MBHs leading to the ejection of the lightest BH \citep[e.g.][]{Hoffman2007}.
Here we briefly visit a case in our simulation which could also lead to an ejected BH on the outskirts of a large galaxy/halo.

Panel (e) illustrates a possible ejection or orbital disruption of MBHs during the quadruple encounter.
In this system, all four MBH masses differ by an order of magnitude at the beginning of the quadruple encounter, which could be a reason for the frequent orbital disruption in this system.
By examining the pairwise separation of the MBHs, we see that the orbit size of the smallest BH increases to above $500\,{\rm kpc}$ soon after the quadruple phase at $z=3$, followed by a sudden drop in its luminosity by almost four orders of magnitudes. 
The second and third largest MBH also experiences orbital disruptions at around $z=2$, leaving three MBHs at more than a hundred kpcs from the galaxy center.

Previous works on triplet-induced MBH ejection mainly focused on the effect of close-separation three-body interactions, but the dynamic and chaotic environments of multiple galaxy mergers may also lead to the ejection or at least the orbital enlargement of MBHs, especially with large mass contrast between all infalling BHs.
The case we show here provides an interesting and compelling case for a galactic-scale ejection mechanism and may be of interest to a more detailed simulation study.

\subsubsection{Mass Ratio as an Indicator of Evolution Paths}

\begin{figure}
\centering
\includegraphics[width=0.48\textwidth]{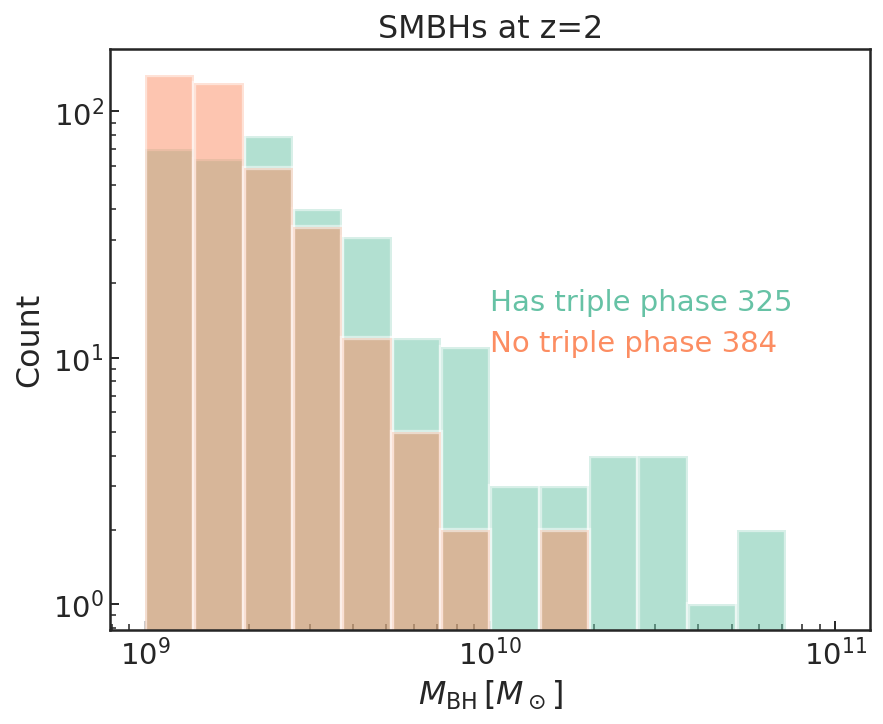}
  \caption{The mass distribution of the $M_{\rm BH}>10^9\,M_\odot$ SMBHs in \texttt{ASTRID} at $z=2$ that has gone through a triple MBH phase between $z=3$ and $z=2$ (\textit{green}) and those that have not (\textit{orange}). Notably, among the 17 heaviest $M_{\rm BH}>10^{10}\,M_\odot$ SMBHs in the simulation, 15 of them have been involved in an MBH triplet in the near past.} 
  \label{fig:smbh}
\end{figure}

In addition to the electromagnetic signatures of the MBH-powered multiple AGN that we have studied earlier, previous works have also shown that triplet MBHs could result in an enhanced MBH merger rate and gravitational wave sources in both the LISA and the PTA bands \citep[e.g.][]{Bonetti2018a,Bonetti2019}.
However, their work is concerned mostly with the $\rm pc$-scale three-body interactions, and less with the large-scale interaction between the multiple galaxies.
As we have seen from the earlier case studies and also in \cite{Chen2022}, the earlier tidal disruption among the host galaxies of tuples may prevent the BHs from forming a binary or close triplet system in the first place, and so the triple-MBH binary formation channel may not take place under those scenarios.
From the case studies, we have found that quick mergers between all three (or all four) MBHs are usually found in equal-mass MBH tuple systems, while larger mass contrasts often result in orbital stalling or disruptions.
In this section, we show all the subsequent MBH mergers among our MBH triplets and their correlation with the mass ratios between the three objects.

In Figure \ref{fig:mergers}, 
we divide the triplet systems by the number of mergers each system has gone through in the next Gyr, and show the mass ratios between the MBHs among each population.
We also show the number of systems that may have at least one MBH ejection or significant increase in orbital size during this time (the ``kicked" MBHs).
To quantitatively define a ``kicked" MBH, we compute the ratio between consecutive apocentric radius between each MBH in the triplet and the central MBH.
If we see a consistent and sudden increase in orbital size by a factor of more than three in at least three full orbits, then we categorize the MBH as a ``kicked" MBH.

Among the close-separation triplets, about half would go through at least one merger among the triplet, out of which only $20\%$ go through two consecutive mergers.
This indicates that even though multiple MBH systems are prevalent among massive halos, the actual chance of forming a close triplet encounter could be low.
About $30\%$ of triplet systems do not merge within the simulation over the next Gyr, out of which half has gone through ``kicks" which leads to an increase in the orbital size.
For the large-separation triplets, the chances of consecutive mergers are lower at $\sim 4\%$, but $20\%$ have gone through at least one merger.

The most noticeable difference between the no-merger and merged triplet systems is the pairwise mass ratio.
The triplets that have all three MBHs merged are dominated by major merges with pairwise mass ratios above $0.1$.
This is both due to the small dynamical friction time in major mergers, and the less severe tidal disruptions of host galaxies.
The triplets that never merged and those that may have experienced an orbital disruption are characterized by a low mass ratio of $\sim 0.03$.

We note that due to our resolution limit, the in-simulation mergers are precursors of bounded binary systems, but if the MBHs in triplet never make it to an in-simulation merger, then the triple-induced MBH mergers are less likely to happen.
However, in the cases where we do observe two or more in-simulation mergers from the MBH tuples, the mergers usually happen relatively quickly (as was shown in panel (c) of Figure \ref{fig:histories}), in which case the third MBH may have an immediate in pact on the orbital evolution of the original MBH binary.

\subsection{MBH Triples and Ultra-Massive Black Hole Formation}
\label{sec:smbh}
Previous work by \cite{Ni2022b} has reported the formation of an extremely massive SMBH with $M_{\rm BH} \sim 10^{11}\,M_\odot$ around $z=2$ in \texttt{ASTRID}, which formed during a high-accretion phase after the consecutive mergers between three $M_{\rm BH} > 10^9\,M_\odot$ SMBHs.
This points to the important role of MBH tuples in the formation of the most massive SMBHs in the Universe.
We have also found in the case studies that one or more of the MBH involved in triplet systems experience rapid mass growth during the triplet encounter, during which their mass increases by more than ten folds.
In this section, we look at the relationship between SMBH formation and triplet encounters.

In Figure \ref{fig:smbh}, we plot the mass distribution of the $M_{\rm BH}>10^9\,M_\odot$ SMBHs in \texttt{ASTRID} at $z=2$ that have gone through a triple MBH phase between $z=3$ and $z=2$. 
Among all $M_{\rm BH}>10^9\,M_\odot$ SMBHs, $50\%$ have recently been involved in an MBH triplet, which would likely contribute to their rapid mass growth both by MBH mergers and by the supply of gas brought in by other MBHs in the triplet.
Notably, among the 17 heaviest $M_{\rm BH}>10^{10}\,M_\odot$ SMBHs in the simulation, 15 of them have been involved in an MBH triplet in the near past.
Between $10^{9.5}\,M_\odot$ and $10^{10}\,M_\odot$, the triplet-involving SMBHs still dominate in numbers.
Most of the $>10^{9}\,M_\odot$ SMBHs that have not been through a triplet phase have masses barely crossing the $10^{9}\,M_\odot$ threshold.
We leave the detailed study on the role of the triplet phase in the formation and growth of $10^{10}\,M_\odot$ SMBHs to future works.

%% file: Sec4_Conclusion.tex
\section{Conclusion}

In this work, we characterize the properties and evolution of triple and quadruple MBH systems at $z=2, 3$ within the \texttt{Astrid} simulation.
We define a tuple system to be one where all MBHs exceed $10^7 M_\odot$ and have pairwise separation $\Delta r < 200\ \rm kpc$.
Our primary results are summarized below:
\begin{itemize}
    \item   At $z=2$ and $z=3$, $\sim 4\%$ of the $M_{\rm BH} > 10^7\,M_\odot$ MBHs are involved in an MBH tuple (at least three MBHs separated by less than $200\,{\rm kpc}$). 
    These systems are only found in halos with $M_{\rm halo} > 10^{12} M_\odot$, and occur in $M_{\rm halo} > 10^{13} M_\odot$ with $>10\%$ likelihood.
    
    \item Among MBH triplets, only $\sim 65\%$ have three BHs simultaneously luminous with $L_X>10^{42}\, {\rm erg/s}$. Such systems are typically found at a maximum separation between $40\,{\rm kpc}$ and $120\,{\rm kpc}$.
    Factoring in the fluctuation in AGN activity, only $60\%$ of the triplets are simultaneously bright for $>80\%$ of the time over the course of a Gyr. The 3-AGN time increases with the triplet separation.
    \item The 3-AGN triplets have masses ranging from $10^7\,M_\odot$ to $10^9\,M_\odot$ and galaxy masses between $10^{10}\,M_\odot$ and $10^{11}\,M_\odot$.
    Most of them are captured before the merger of all three galaxies.
    
    \item MBH tuples can lead to both quick infall time and mergers (in $5\%$ triplets all three MBHs would merge within a Gyr, and $15\%$ go through one merger), and wandering MBHs due to tidal disruption. We also found possible cases of triple-induced runaway MBH in 
    $5\%$ of the triplets, where at least one MBH is ejected to a large distance from the halo center during the dynamic interactions within the tuples.
    
    \item The subsequent in-simulation mergers between MBH triplets are correlated with their mass ratios. Systems with all three MBHs merged involve similar mass MBHs with pairwise mass ratios above 0.1. Stalled orbits or runaway BHs are more frequent in systems with $q\sim 0.01$.
    \item The heaviest MBHs among in tuples experience a phase of rapid mass growth, during which their mass increases by more than ten folds following the tuple formation, both by efficient gas accretion and by BH mergers. 
    Among all the SMBHs with $M_{\rm BH} > 10^9\,M_\odot$ at $z=2$, half were involved in an MBH tuple between $z=2$ and $z=3$. 
    Out of the 17 $M_{\rm BH} > 10^{10}\,M_\odot$ ultra-massive BHs at $z=2$, 15 of them have been through an MBH tuple phase in the past $\sim 1\,{\rm Gyr}$.

\end{itemize}

While MBH tuples are rare, we do anticipate their existence in a majority of massive halos.
Our results point to the likely important role of MBH tuples in producing runaway MBHs and some of the heaviest SMBHs.
We also characterize the MBH tuples that are most likely found in the rapid binary formation which can lead to enhanced LISA rates.
Finally, we want to point out that in addition to the multiple AGN systems and the GW emissions, MBH tuples and the runaway MBHs associated with them can also leads to flares in stellar tidal disruption events \citep[e.g.][]{ChenXian, Stone2012}, which is potentially observable at a rate of a few per year by the  Vera C. Rubin Observatory.
More detailed studies and high-resolution simulations along any of these directions could lead to further insights into how MBHs behave under the extreme environment.